# Mechanical and Thermal Tests of the Containers of Liquid Mirrors.

E. F. Borra[2], R. Content[1], G. Tremblay[2], A. Daigle[2], Y. Huot[2]

SEND ALL EDITORIAL  CORRESPONDANCE TO R. CONTENT

[1] University of Durham, Astronomical Instrumentation Group, Department of Physics, South Road, Durham, DH1 3LE, UK

[2] Département de Physique, Génie Physique et Optique, COPL, Université Laval, Canada G1K 7P4

Subject headings: Telescopes, Instrumentation, Liquid Mirrors

**Abstract**

We give a generic description of a liquid mirror system and summarize some practical information useful to making them. We compare laboratory measurements of deformations, due to the weight of mercury on the container of a 3.7-m mirror and to temperature changes on a 1-m container, to finite element computer simulations. We find that the measurements agree reasonably well with the numerical simulations. The measurements on the 1-m container show no evidence of fatigue after a few thermal cycles. These results validate the computer models of containers described in a companion article.

## 1. Introduction

The liquid mirror technology has now been fully demonstrated in the laboratory and in observatory settings for 4-m class liquid mirror telescopes (LMTs). Optical tests have shown that a 1.5-m diameter liquid mirror (LM) was diffraction-limited (Borra et al. 1992, hereafter referred to as Paper I). Paper I gives a wealth of information on the basic LM technology. Tests of a 2.5-m diameter liquid mirror (Borra, Content, & Girard 1993; Girard & Borra 1997, hereafter referred to as Paper II) also showed diffraction limited performance. Optical tests of a 3.7-m LM (Tremblay & Borra 2000, hereafter referred to as Paper III) lead us to the conclusion that mirrors in the 6-m class will work, not excluding even larger mirrors.

Liquid mirrors are interesting in several areas of science. For example, the University of Western Ontario has built a lidar facility that houses a 2.65-m diameter liquid mirror as receiver (Sica et al. 1995). A lidar facility has also been built and operated by the University of California at Los Angeles (Wuerker 1997). The NODO orbital debris observatory run by NASA ( Hickson & Mulrooney 1997) has demonstrated that the technology, although in its infancy, is sufficiently robust to be used for extended periods of time in astronomical observatory settings. The 3-m NODO LMT is extensively discussed by Mulrooney (2000) who describes its operation and performance. While the UWO and UCLA LMTs are used for the atmospheric sciences and are not imagers (but operate in very harsh climates), the NODO telescope is of particular importance to astronomy since it is an imager that has continuously operated for several years and is beginning to yield published astronomical research (Hickson & Mulrooney 1997, Cabanac & Borra 1998). Liquid mirrors are also used as reference surfaces to test conventional optics (Ninane & Jamar 1997). They have been used to test the auxiliary optics of the auxiliary 1.8-m telescopes of the VLT. An International 4-m LMT (ILMT) has been mostly funded is in an advanced stage of planning and construction should soon begin (J. Surdej, private communication).

Over the past several years we have carried out a number of unpublished engineering studies at Laval University that investigated the mechanical characteristics of containers, bearings and mounts. The earlier studies involved analytical models that were later superseded by numerical computations carried out with commercial finite-elements software. The results were presented in the form of internal reports (in French). The unpublished M.Sc. thesis of Arrien (1992), also in French, is the only document that is readily accessible. It explores with finite element computations the limits of containers of liquid mirrors made with composite materials. The main conclusions reached by Arrien (1992) are that composite containers can be useful up to diameters as large as 6 meters but that they become too heavy and expensive for significantly larger diameters. Hickson, Gibson & Hogg (1993) have studied, analytically, the properties of composite containers for liquid mirrors. They give a number of useful analytical expressions. They also describe an improved design for a composite container and discuss construction techniques as well as materials.

Building on the previous studies, we have made detailed engineering studies of the containers. The previous designs had some limitations that make them difficult to scale to larger sizes. The improved design is lighter, more resistant and more rigid. A companion paper (Content 2003) carries out finite element computations of containers, based on this design, under load. It also evaluates the important temperature effects on the containers. The goal of the two articles is to produce practical recipes that will allow one to build a working LMT container without resorting to his own finite element computations. In the present article, we give the results of a series of measurements made on a 3.7-m container with 3.6-m clear aperture F/1.2 and on a 1-m flat container, comparing them with finite elements computer simulations (Content, 2003).

## 2. Generic description of a liquid mirror system

The main components of a liquid mirror are shown in Figure 1. The container is the most conspicuous one. Its function is obvious: It holds the reflecting liquid. Although, at first sight, it may seem like an unsophisticated item, in reality, one must pay close attention to its design and care during construction. The container must be stiff to hold, without excessive flexures, the heavy reflecting liquid metal, yet be light since high-precision bearings are expensive and their cost increases rapidly with the maximum load that they can bear. Paper I discusses at length a variety of issues related to the container and mirror.

Containers can be made with a variety of construction techniques. Our early containers were made of a flat plywood disk bolted on an aluminum plate. A metal strip was bolted around the circumference of the disk. This simple construction technique is adequate for small mirrors having diameters of the order of 1 meter, but it is not good enough for larger ones. The container of the 1.5-m mirror tested in paper I used a similar design but the flat disk was made of fiberglass laminated over foam. The flat disk was parabolized by spincasting polymer resins. This design was adequate for this mirror but would have produced too heavy a container for larger mirrors. The container of the 2.5-m (Paper II) as well as those of all subsequent containers are made of Kevlar laminated over a foam core. The original design based on the analytical solutions in Hickson, Gibson & Hogg (1993) was investigated with theoretical stability calculations (Content 1992) that indicated that it was too weak and would flex and spill its contents. The container was therefore strengthened, on the basis of the finite element computations (Arrien 1992), by adding several plies of Kevlar in its central region The deflections measured for the modified container under azimuthaly uniform load were found to be in good agreement with the finite element computations (Arrien 1992, Girard 1996).

The present article discusses containers built with composite materials (Kevlar laminated over a foam core). One can distinguish several components in this type of container. Figure 2 shows a cut-off of the generic container of a liquid mirror and identifies its main components. The most visible part is the underlying support structure that rests on a flange that mates it to the bearing. The support structure holds the top surface which plays a most important role since it makes contact with the reflecting liquid. In practice, it is necessary to work with thin layers of liquid. Thin layers are needed to minimize weight and, most importantly, to dampen all disturbances transmitted to the liquid surface. <u>One cannot overemphasize the importance of working with thin layers</u>: Our experiments (Paper I, Paper II, Paper III) show that mercury mirrors with layers thicker than 2 mm are overly sensitive to disturbances. It is best to work with a layer of mercury of the order of 1-mm thickness or less. Paper I describes how we make thin mercury layers that would normally be unstable. Obviously, the top surface must follow a parabola to a fraction of the thickness of mercury. Spincasting a polymer resin gives a convenient way to accomplish this. First, the top surface of the support structure must be parabolized to within a few mm. The liquid polymer resin is then poured on the rotating container, acquiring a parabolic shape. Finally, it is allowed to harden, giving a solid parabola.

Spincasting is not a trivial operation for it requires a good knowledge of the commercially available resins that must fulfill several criteria. The resin must polymerize at room temperature. Its viscosity must be sufficiently low (a few hundred centipoises) to allow it to flow easily and follow the parabolic equipotential within the time it remains liquid. Polymerization times at room temperature must be large enough (of the order of one hour) to allow the resin enough time to follow an equipotential. On the other hand, it must be short enough to be practical. Note that polymerization times are very temperature sensitive. Resins can heat up considerably upon polymerization. The temperature they reach increases rapidly with volume and thickness of resin. Early in our experiments, we had a spectacular failure with an epoxy resin that was poured in a 2-cm thick layer: It boiled violently before our eyes. Physical properties, handling and safety features can vary widely among resins. Resins shrink upon polymerization. Our earlier containers were spuncast with polyester or epoxy resins but we found that they behaved like bimetallic plates and warped with temperature variations (Paper I). We now spincast the top surface of our containers with a soft polyurethane resin, having a much lower elastic coefficient (Young modulus ) than epoxy, that renders the effect negligible. Polyurethane has a secondary advantage over epoxy: mercury sticks better to a urethane surface than to an epoxy one. This advantage is of interest because it facilitates closing the mercury surface during startup (Paper I).

The mercury layer occasionally breaks locally upon startup with a subcritical layer thickness (< 4mm). With an epoxy top surface, a local break in the mercury surface grows rapidly and destroys the layer, forcing one to restart the mirror. This does not happen with a polyurethane surface for mercury sticks to it, usually giving one enough time to repair the local break by coaxing manuallly the mercury to close around the break by stroking it with a hard surface.

The coning error of an airbearing and metal base increases if the mass distribution is not axially symmetric. Since the mass distribution of a container is unlikely to be perfectly symmetric, it is necessary to balance it. Static balancing is easily done by first tilting the axis of rotation of the mirror, causing the unbalanced container to rotate. After a few damped rotations around the axes of rotation, it stops at the position angle of the heaviest part of the container. Small weights are then added, by trial and error, until the container no longer has a heavier side. Static balancing is however insufficient for the thickness of the mercury layer is not uniform, due to flexures and imperfections of the resin top surface. Dynamical balancing is needed, at least with mirrors that use airbearings, albeit it may not be needed with the much stiffer oilbearings. It is performed by trial and error on the rotating mirror, adding weights until the coning error, measured by looking at the movement of the PSF in a full mirror rotation, decreases below detection. This is a lengthy procedure which, fortunately, is done once and for all; unless the container is resurfaced. However, we found that dynamical balancing varies with the thickness of mercury, as should be expected.

## 3. Laboratory tests of containers

Two containers were designed, using the ANSYS program described in Content (2003), for the Liquid Mirror Laboratory of Université Laval and built there: A 3.7-m (3.6-m clear aperture F/1.2) container, which has since been thoroughly tested optically (Paper III), and a 1-m flat container designed for studying the effects of temperature changes. The mechanical and thermal behaviors of these containers where thoroughly tested, compared to the final element computations, and give important tests of the reliability of the finite element computations. The principal parameters, described in more details by Content (2003) are listed in Table 1.

## 3.1. Mechanical tests of the 3.7-meter container under load

Because the 3-7-m mirror is used in the laboratory and is not subject to temperature changes, the design was made to maximize the rigidity and minimize weight. In this case, the maximum stresses are well below the largest safe values. The main difference of its design with respect to the containers described in Content (2003), is that there is no aluminum plate on the top of the cylinder. This plate is added to the standard design to reduce the maximum stresses due to temperature changes, something not present in the laboratory.

### 3.1.1. Finite element analysis

A two-dimensional axisymmetric model of the container was written with the ANSYS program of finite element analysis version 5.0 from Swamson Analysis System, using azimuthal symmetry to avoid the need of a three dimension model (Fig. 3). The model is similar to the one described in Content (2003) but simpler. The Kevlar and the cylinder are modelled with skin elements which have one less dimension, using the thickness as a parameter. There also are fewer elements because the deformations, not the constraints, were dominant here since there were no temperature variations. This model was straightforward to use to simulate a uniform layer of mercury, which is also azimuthally symmetrical, but not to simulate bottles of mercury at one or 2 positions. It is however possible to use the program with non-axisymmetric loads by decomposing them in a Fourier series in the azimuthal direction (Content 2003). We used 63 terms to properly simulate the load, higher order terms were very small and the series converged rapidly.

### 3.1.2. Deformation under uniform load.

We carried out mechanical tests that consisted in measuring the deformation at different points of the edges of the container, after placing bottles of mercury on the top surface. The bottles of mercury were uniformly dispersed on the surface to simulate layers of mercury having thicknesses varying from 0 to 3.0 mm in steps of 0.5 mm and the deformations were measured at 5 positions on the container edge. The differences between the deformations measured and the theoretical values are plotted in Figure 4. While individual measurements do not fit perfectly with the calculations, the average of the 5 measurements is well within the uncertainties. The sinus-like wave seen in Figure 4 suggests that a small tilt has been introduced. This is possible since any small decentering of the mercury bottles would create a moment that would tilt the container. Although more measurements would be necessary to verify this explanation, the measurements with loads at a constant azimuth seems to confirm this theory (section. 3.1.3).

### 3.1.3. Deformations due to loads at a constant azimuth

A series of measurements were done after placing bottles of one litre of mercury at two opposite positions, at equal distance from the center and measuring the vertical displacement of the container near the rim on one side. A first bottle was placed on one side, the second on the opposite side, the third again on the first side and so on, adding alternatively a bottle at one of the opposite positions between measurements. Measurements were taken at a radius of 1.82 m with bottles placed at 1.00, 1.70 and 1.79 m from the center. They were also made along 3 other azimuthal directions with bottles placed at 1.79 m. The load can be divided into 2 components, one with an equal load at each of the 2 positions and one with a load on one side only. The former generates a saddle shape deformation while the latter mostly induces a tilt of the container. The total deformation can then also be divided into the corresponding deformations:

$$D = w(n_s d_s + n_a d_a) \qquad (1)$$

where $D$ is the measured vertical displacement, $w$ the weight of one bottle of mercury, $n_s$ (= 0, 1, 2 or 3) the number of mercury bottles on one position for the part of the load equal on each opposite position, $d_s$ the vertical displacement when $1\ N$ is applied at each of the opposite positions, $n_a$ (= 0 or 1) the number of mercury bottles on the position on the measurement side for the part of the load on one side only, and $d_a$ the vertical displacement when $1\ N$ is applied on the position on the measurement side.

The values and standard errors of $d_s$ and $d_a$ could in principle be calculated by taking a least square fit through the measurements. However, the deformation was not exactly the same for each bottle because they were not one over the other but side by side. Also, the measurements showed significant differences with the theory for asymmetric loading. We used a more precise formula to test our theory:

$$D = C\, d_t + n_a r \qquad (2)$$

where $D$ and $n_a$ are as in (1), $C$ is a constant, $d_t$ the theoretical deformation and $r$ the residual when there is one more bottle on the measurement side. $C$ is then only determined by the symmetric loads. If the theory fit perfectly, $C$ would be compatible with $1.0$ and $r$ with $0.0$ considering the standard errors.

The values with bottles at 1.70 m were obtained in a slightly different manner. A fourth bottle of mercury placed at 1.70 m was added to 3 bottles at 1.79 m (because there was no more space). The values were extracted from the data by extending equation (2) to 4 unknowns (2 per radius) instead of 2. This explains why the standard errors are so large at 1.70 m (figures 5 top and 5 bottom: it is due to the small number of measurements.

### 3.1.3.1. Saddle shape deformations

The vertical displacement at 1.82 m, calculated from the measurements, is compared in Figure 5 (top) to the theoretical value calculated with the ANSYS program for symmetrical loads. It shows the ratio of the measured to theoretical values when a load is applied at 2 opposite positions at the radius indicated on the horizontal axis ($C$ in Eq. 2). The resulting values are about 14% larger than the theoretical values. This is not surprising, considering how difficult it is to simulate a system containing Kevlar and foam and to measure the small displacements predicted. For example, the large standard error at a radius of 1 m comes from the very small value of the displacement expected: 16 microns per couple of mercury bottles. At 1.79 m, the differences between the different directions cannot be explained by the standard errors. This suggests that there is a small difference in rigidity between different directions, a result that is not surprising when considering that a layer of Kevlar has a much higher rigidity in 2 perpendicular directions. It is impossible to built a perfectly isotropic container. However, the variations with azimuth are much smaller than for the test with a uniform load (section 3.1.2), that is a standard deviation of 7 % instead of 36 %. This seems to confirm that the uniform load variations may be the result of an unbalanced load.

### 3.1.3.2. Deformations in tilt

The vertical displacement at 1.82 m calculated from the measurements for a load on one side only is compared with the values obtained with ANSYS in Figure 5 bottom. In this case, it is the difference, not the ratio, between measured and theoretical values that appears when a bottle of mercury is placed at the indicated radius ($r$ in Eq. 2). The residuals are far larger than can be explained by the standard errors. Since we know from the other tests that the container behaves as predicted under a balanced load, we conclude that the central part does not behave as predicted when an unbalanced load is applied while the rest of the container does. The residual is then the result of an unexpected phenomenon in the center. This is surprising since the central part is made of aluminium glued to a concrete floor. However, if the problem comes from the fact that the rigidity in the center is smaller than expected, the residual would be proportional to the applied torque. This is not the case as can be seen on Figure 5 bottom for the data points of the only direction where more than one torque value was applied (white circles). The origin is out of the 95% confidence interval calculated with the standard errors of the 3 data points.

The data points show that the central part is less rigid when the container is perfectly balanced and increases in rigidity with the torque of the applied load. This is typical of many mechanical systems that need a minimum load to become fully rigid. It would help to explain the residual tilt of the test with a simulated 1 mm layer of mercury (section 3.1.2) since this could be obtained with a smaller unbalanced load than expected. In our case, this could be the result of screws that were not sufficiently tight. For the measurements, the container was attached with only four screws to a plate glued to the floor. By increasing the applied torque, a larger part of the screws should touch the metal and the rigidity should increase. Assuming that the residuals follow an exponential curve through the origin and using the other measurements, the limit with the load at an infinite radius (giving an infinite torque) is 67 microns. This means that the vertical displacement at 1.82 m is increased by up to 67 microns due to this lack of rigidity. This corresponds to only 6 microns at the radius of the screws which makes them the prime suspects.

Not only is 67 microns very small compared to the thickness of mercury that we use, but the design of future containers (Content 2003) should include a proper rigid center and attachment to its base. There are, however, many uncertainties in the measurements so we cannot conclusively prove our case. If we simply do a linear regression of the measured deformations as a function of the theoretical deformations, we would conclude from the slope that our angular rigidity is (0.72 +/- 0.10) of the theoretical value with a 95% confidence interval. Using only the measurements at a radius of 1.79-m would give a value of 0.50, but in both case we ignore informations telling us that this is too pessimistic. We therefore conclude that our measured angular rigidity is equal to (0.8 +/- 0.3) of the theoretical value.

To resolve this question, we tried to measure the rigidity using a different approach. The lowest resonance frequency of the container is the angular vibration in tilt. If this frequency and the container inertia are known, the angular rigidity can be calculated (Tremblay 1999). The angular inertia in tilt was evaluated by measuring the rotational inertia and assumed the two were in the same proportion than for a disk, then corrected for the distance between the center of gravity and the point of attachment. The vibrational amplitude was measured when submitted to mechanical excitation at different frequencies. The lowest resonant frequency was obtained from these measurements. The result is also an angular rigidity of (0.8 + - 0.3) of the theoretical value.

### 3.1.4. Weight of the container

We measured the weight of the bare container, without spincasting resin and mercury. It is significantly heavier than expected: 195 kg instead of 138 kg. The discrepancy seems to come mostly from excess resin used to laminate the Kevlar layers and resin thickened with microspheres used to improve the surface of the foam before laminating the Kevlar. Measurements show that the laminated Kevlar layers are thicker than expected by at least 30%. While this does not change the mechanical properties because the additional resin has a negligible effect on the mechanical strength and the rigidity of the container, it increases the weight. We laminated the Kevlar layers by hand, using a roller. It is likely that better techniques (e.g. vacuum lamination) will yield more uniform layers and a weight in better agreement with the values quoted in Content (2003). However, it would be wise to assume a container heavier than suggested by the finite elements when designing the mirror because a heavier container has implications in the choice of the components (e.g. the bearing). Note that the measured weight of the 1-m container (see below), which was made of foam and Kevlar with a steel base and included a better evaluation of the Kevlar and resin weight, fitted well with the theoretical value. Applied to the 3.7-m, it does not completely explain the value of 195 kg. However, the discrepancy is only a few percent of the *total* weight of the container plus polyurethane resin plus mercury.

### 3.2. Tests of a 1-m container subjected to thermal changes

Because astronomical mirrors operate at the ambient temperature of the observatory site, they undergo important temperature variations. Because the mirror must operate with a mercury layer having a thickness of 1 mm or less, the surface of the container must hold its parabolic shape to a fraction of a millimeter. Composite containers mix materials having different thermal expansion coefficients and should be expected to shift shape as the ambient temperature changes. It is thus necessary to understand the behavior of the container with respect to temperature variations. This has been done with finite element computations as reported in Content (2003). In this section, we discuss laboratory studies of a small container and compare the theoretical temperature effects to those obtained on the container subjected to temperature variations. The comparison between the two can be used to validate the finite element computations in Content (2003).

### 3.2.1. Description of the container and tests

The 3.7-m container is large and we did not have the large cold room needed to cool it. We therefore designed a 1-m container with the numerical techniques described in Content (2003), built it and placed it in a cold room where the temperature can be varied between the ambient temperature and -30°C. A small-scale model of the larger containers would have given small deformations that would have been difficult to measure; therefore, to maximize the deformations of the 1-m container, the central thickness was made proportionally larger than those of the other containers (39% of diameter instead of 14% for the 3.7-m), and the aluminum cylinder was replaced by a Kevlar cylinder because of the larger difference between the thermal expansion coefficients of Kevlar and foam. Over several months, eight cycles of lowering and increasing the temperature were carried out. For each cycle, we performed measurements of deformations at the minimum and maximum temperature (sometimes at additional values) giving a total of 25 series of measurements, themselves split in 4 groups (F1 to F3 and B1). For each series, the deformation was measured at 28 to 33 points (depending on the series) on the surface of the flat container. A total of 749 measurements were taken. The temperature inside the room was measured with a thermometer.

The measurements were taken between specific points, defined by small spheres glued on the surface of the container, and the face of an aluminum table that stood supported by steel columns over the face of the container. During the first group of measurements (F1), while the container itself has roughly performed as expected, a problem occurred with the base on which the container was bolted. It was made of a square of plywood attached to the container by a central screw. At low temperature, the plywood was twisted and forced in the vertical direction. As a consequence the container was tilted and raised in an unpredictable manner. To minimize this problem, a model of the deformations was fitted to calculate the height and tilt of each series. They were then subtracted from all the measurements of the series. However, the data from F1, especially the piston and tilt terms, are less reliable than the data from F2 , F3 and B1. For those groups of measurements, the plywood base was removed and replaced by a steel ring glued to the container. This gave smaller piston and tilt variations.

The temperature of the cold chamber was repeatedly cycled from maxima of 8 to 23 degrees C to minima between -10 and -30 degrees C. During each cycle, the temperature was stabilized at a given temperature and measurements were taken. Needless to say, this took a considerable amount of time. The measurements were taken over several summers. The B1 group is particularly interesting, for we flipped the table, using thus its former back surface as the new reference surface. We did this because we suspected that our reference table warped slightly with temperature. We indeed found that the B1 measures behave somewhat differently from the other 3 (e.g. see Fig. 6 and 7).

As for the 3.7-m, the container was modeled with the ANSYS program of finite element analysis (Fig. 8). It was used to calculate the theoretical deformation as a function of radius for a variation of 1 °C.

### 3.2.2. Data  analysis method

The measurements were difficult as the deformations were small, so the precision is lower than we hoped for. Only by using the proper statistical method over all the series of measurements could we obtain usable results. Each of the 4 groups was analyzed separately since the container and in some case the table were moved between measurements, then the 4 values were averaged giving to each group a proper weight. Two kinds of averaging were done: in one case the deformation per degree C (in fact per drop of 1 °C) for each of the 33 points of measurement on the container was calculated from the 25 measurements at different temperatures. This was compared with the theoretical deformation obtained by finite element analysis. On the other, a temperature was calculated for each of the 25 series from the measured deformations at the 33 points using the theoretical deformation per degree C as the reference. This was compared to the temperature of the thermometer. These calculations were done for 3 types of deformations: the surface deformations which are relative deformations, in this case using the radius of largest theoretical deformation as the reference (19 cm); the global contraction which is the deformation at that reference radius and is a trivial piston term that can readily be compensated by moving the detector up and down; and the total deformation which is the sum of the two.

Simple calculations could not be done because to calculate the average deformation at each of the 33 points of measurement it was first necessary to remove the random tilt that appear at each temperature especially for the F1 group which has an unstable plywood base. This needed to fit a model at each temperature but it could not be done without first calculating the reference position at 0°C for each of the 33 points. Only by fitting a model through all the 33 points at all the temperatures of each of the 4 groups was it possible to calculate tilt and reference 0°C at the same time. The equation used was

$$H(x,y,T_t) = H_0 + T_d(T_t)H_a(\sqrt{x^2 + y^2}) + H_{r0}(T_t) + H(x,y,0) + x\theta_x(T_t) + y\theta_y(T_t) \qquad (3)$$

with the conditions:

$$H_a(r_0) = T_d(0) = H_{r0}(0) = \sum H(x,y,0) = \theta_x(0) = \theta_y(0) = 0 \,, \tag{4}$$

where $H$ is the distance between the surface and the table, $x$ and $y$ are the coordinates on the container surface, $T_i$ the thermometer temperature, $H_0$ the average distance between surface and table at 0°C, $T_d$ the temperature calculated from the surface deformations, $H_a$ the theoretical surface deformation per degree C calculated with the ANSYS program as a function of radius, $H_{r0}$ the deformation relative to the table at radius $r0$, $\theta_x$ and $\theta_y$ the angular tilt in $x$ and $y$, and $r0$ the reference radius for the global contraction and the surface deformations ($r0 = 19$ cm). Functions of $T_i$ and $H(x,y,0)$ are modeled as a series of constants, one per respectively temperature and point of measurement. The largest fit (group F1) involved 368 equations and 81 unknowns. Equations 3 and 4 must be slightly modified when calculating $T_d$ from the total deformation as in figure 6: the $H_{r0}$ term must be removed as the condition on $H_a$, which becomes the theoretical total deformation.

The global contraction is obtained from the slope of $H_{r0}$, then corrected for the contraction of the feet of the table. The model is then described by

$$H_{r0} = (G_c + \alpha L)T_t + const.,$$  (5)

where $G_c$ is the global contraction per drop of one 1°C, $\alpha$ the thermal expansion coefficient of carbon steel (average of 7 values from different sources), $L$ the length of the feet of the table, and the constant is in principle equal to 0.

The surface deformations were obtained for each point of measurement from the slope of the residual deformation (after the surface tilt and the reference deformation [global contraction] were removed at each temperature) as a function of temperature. The model then gives

$$H(x_i, y_i, T_t) - H_{r0}(T_t) - x\theta_x(T_t) - y\theta_y(T_t) = A_i T_t + B_i,$$  (6)

where $i$ is the order number of the point of measurement (1 to 33), $A_i$ is the surface deformation per degree C, and $B_i$ is a constant.

The 4 series were averaged using their standard errors to calculate their proper weight, then the weight of series F1 was reduced because of its unstable base, and the weight of series B1 was increased because it is the only one with the table inverted and should then have a weight nearer to 1/2. For the surface deformations, we have

$$Ftot = (1/3)\ F1 + (1/9)\ F2 + (2/9)\ F3 + (1/3)\ B1$$  (7)

and for the global contraction, where the weight of $F1 = 0$ due to the unreliable plywood base,

$$Ftot = (1/6)\ F2 + (1/3)\ F3 + (1/2)\ B1.$$  (8)

Where Ftot is a deformation or a temperature, and *F1, F2, F3* and *B1* are the deformations or temperature of the corresponding group.

### 3.2.3. Results of the temperature tests

In Figure 6, we plot the best-fit temperature obtained from our data to the actual temperature of the cold chamber at the time of the measurements. We see that there is a good agreement. The less reliable F1 measurements are not plotted in the figure because of the aforementioned problems with the plywood base. Although noisier, the F1 measurements are still in reasonable agreement with the theory. Figure 7 gives the ratios between the experimental and theoretical deformations as a function of group. We only give the surface deformation for the F1 group, the global contraction being too unreliable even if it gives (by chance?) a value compatible with the other groups. Figure 7 shows that the ratios of the surface deformations are <1, with the exception of the B1 group, indicating that the experimental values are smaller than the theoretical ones. A possible explanation is that the difference between the CTE coefficients of Kevlar and foam are smaller than expected. Since the theoretical and experimental values of the global contraction (which is dominated by the contraction of the foam) agree well, there is good agreement between the theoretical and experimental CTE of the foam and it is the CTE of the Kevlar that carries most of the error. The CTE of Kevlar is very small; however, since the Kevlar threads are arranged in a rectangular grid, the CTE at 45 degrees to the threads is much larger and depends on the quantity of epoxy glue actually used. The anisotropy is minimized by using several layers of Kevlar cloth, with the threads oriented at appropriate angles. However, the finite number of layers and the difficulty to align them homogeneously on the highly conical bottom inevitably leads to some residual anisotropy that can explain the smaller surface deformations.

The difference between the F groups and B1 suggests that the table slightly changes shape with temperature. The difference of the surface deformations would be explained by a bimetallic deformation where the top and bottom surface of the table have a CTE difference of about 0.5 %. This is why we increased the weight of B1 when averaging, to partly cancel this effect.

Figure 9 gives the deformations of the 25 series, normalized at a 1°C temperature drop, predicted and measured for each point. The agreement is good, with the exception of points 7 to 11. These points are right above the Kevlar central cylinder so that the discrepancy may be due to a CTE slightly larger than assumed for the Kevlar-epoxy mix. Also, it is difficult to have the model and the container to fit perfectly. The deformation of a modified model is shown in Figure 9. The edges of the cylinder were rounded and the CTE of Kevlar slightly increased. The result is nearer to the measurements and gives an idea of the uncertainty of the modeling.

Figure 10 shows the same average experimental and theoretical deformations, normalized a 1°C temperature drop, as function of radius for the 33 measurement points. Some radius values have more points than others (e.g. 45 cm) because of the way the points were distributed throughout the surface. The agreement is good, except at the center where the Kevlar cylinder is located.

Another subject that can be studied from our data is the possibility of deformations not predicted by the theory, asymmetric deformations for example. For this, we looked at our calculated tilts but found no relation with temperature. On the contrary, the 8 distributions (one $\theta_x$ and one $\theta_y$ per group) are together consistent with random noise (7 out of 8 slopes inside a 90% confidence interval).

In summary, the values of the global contraction and global deformation are in reasonable agreement with the theory. The surface deformations are smaller than expected, but they also have a larger uncertainty than the global contraction and global deformation.

What lessons can we draw for the larger containers? The greatest deviations were found above the central cylinder at the center of the container. This shows the importance of using a cylinder with well-known CTE and shape for correct modeling. An aluminum cylinder would have more reliable CTE and shape. We think that the anisotropy of the Kevlar skin is responsible for some of the disagreement between theory and experiments. This problem should be less severe for the larger containers because there are more layers of Kevlar, allowing for a better isotropy since the layers are laminated with the fibers at an angle with respect to each other. This is especially important for the central part where the thermal deformations have a larger effect. Finally, the angle of the cone is significantly smaller for the larger containers, allowing for an easier angular positioning of the Kevlar threads, therefore allowing more isotropic properties.

Another reason for making the measurements on the 1-m was to verify if there are some permanent deformations that build up with successive cyclings of the temperature. A comparison of the calculated and measured surface deformations at ambient temperatures (Fig. 11) does not show any change that cannot be explained by the measurement errors since the slope of value 0.0 is well inside the 95% confidence interval. We can then conclude that our data show no evidence of permanent deformations, but the errors are too large to argue that their absence is proved. The agreement is excellent for the global contraction, which is dominated by the expansion coefficient of the polyurethane foam (Fig. 7).

## 4. Conclusion

We have carried out mechanical and thermal tests of containers of liquid mirrors designed with the same software and techniques described by Content (2003). The measurements on the 3.7-m are in good agreement with the model of the container under load due to the weight of mercury. Only the measurements in tilt show a difference that can be explained by the poor bonding of the container attached to the ground, something that would not happen for a container placed on a bearing. The measurements on the 1-m flat container are also in good agreement with the model under loads due in this case to a drop in temperature. The differences between experiment and model suggest that the thermal expansion coefficient of the Kevlar may be smaller than assumed. The measurements do not show any evidence of material fatigue after several cycles of reduction and increase of the temperature nor asymmetric deformations. We conclude that we can be reasonably confident that the designs described by Content (2003) will perform as expected.

Composite containers are not necessarily the best design to adopt. The space frame design used by Hickson et al. (1998) is a better one, especially for mirrors having diameters larger than 6 meters. However, composite containers have the important advantage that they are easier to build since they can be constructes with limited technical resources and unskilled inexpensive labor (e.g. graduate students). They are thus particularly interesting in institutions where funds are limited (e.g. universities). This is a major consideration, for liquid mirrors only make sense if they are considerably less expensive than conventional glass mirrors.

## 5. Acknowledgments

This work was supported by Durham University and the Particle Physic & Astronomy Research Council of the United Kingdom and the Natural Sciences and Engineering Research

Council of Canada. We wish to thank F. Arrien and C. Gosselin for early work on containers. A large number of undergraduate engineering students participated in early simulation work: we thank them all.

Table 1:
Principal characteristics of the containers tested

========================================================

| | 3.7-m | 1-m |
|---|---|---|
| Focal ratio | 1.2 | Infinite |
| Material of cylinder | Al | Kevlar |
| Diameter (m) | | |
|     Container | 3.70 | 0.96 |
|     Mercury | 3.65 | - |
|     Clear aperture | 3.60 | - |
|     Cylinder | 0.305 | 0.22 |
|     Aluminum plates | 0.466 | - |
| Thickness | | |
|     Center of container (m) | 0.531 | 0.37 |
|     Cylinder (m) | 0.0127 | 0.0035 |
|     Mercury (mm) | 1.0 to 2.5 | - |
| Weight (kg) | | |
|     Container | 203 | 12.3 |
|     Mercury (2.0 mm + groove) | 295 | - |
|     Spincast resin (0.6 cm) | 69 | - |
|     Total | 567 | 12.3 |

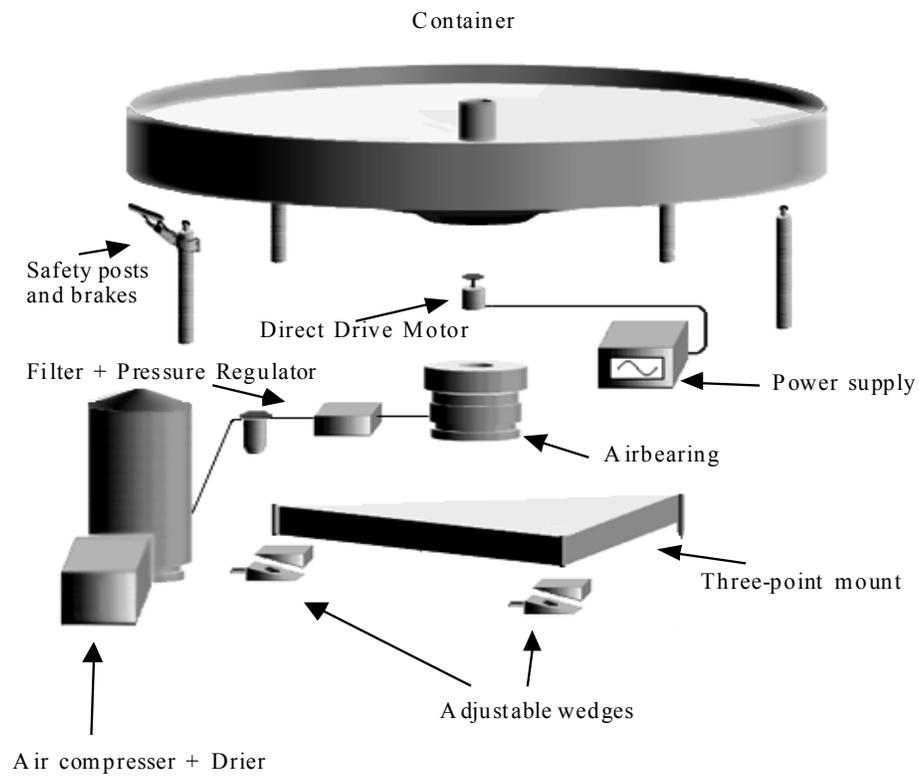

Container

Safety posts
and brakes

Direct Drive Motor

Power supply

Filter + Pressure Regulator

Airbearing

Three-point mount

Adjustable wedges

Air compresser + Drier

**Figure 1:** Exploded view of the basic mirror set-up.

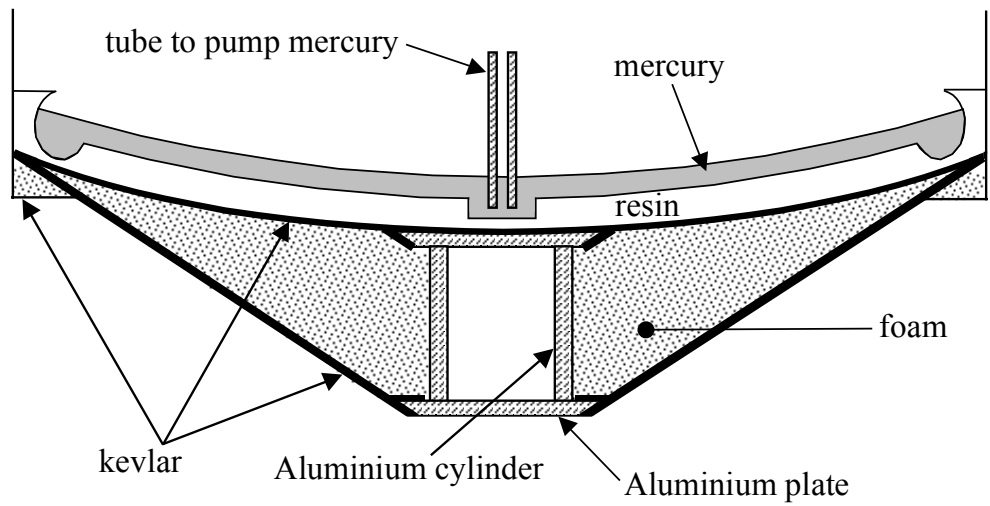

**Figure 2:** Layout of the container and liquid mirror

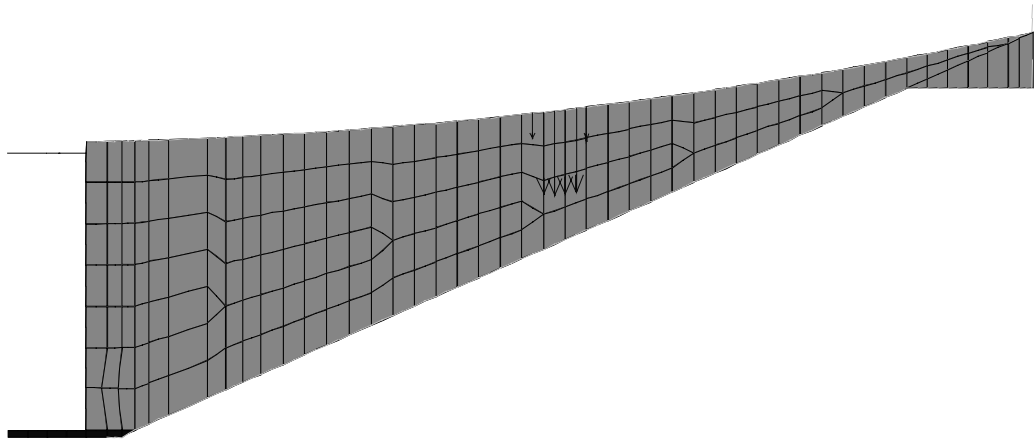

**Figure 3:** Finite element model of the 3.7-m container. The arrows in the center represent the load of a one litre bottle of mercury.

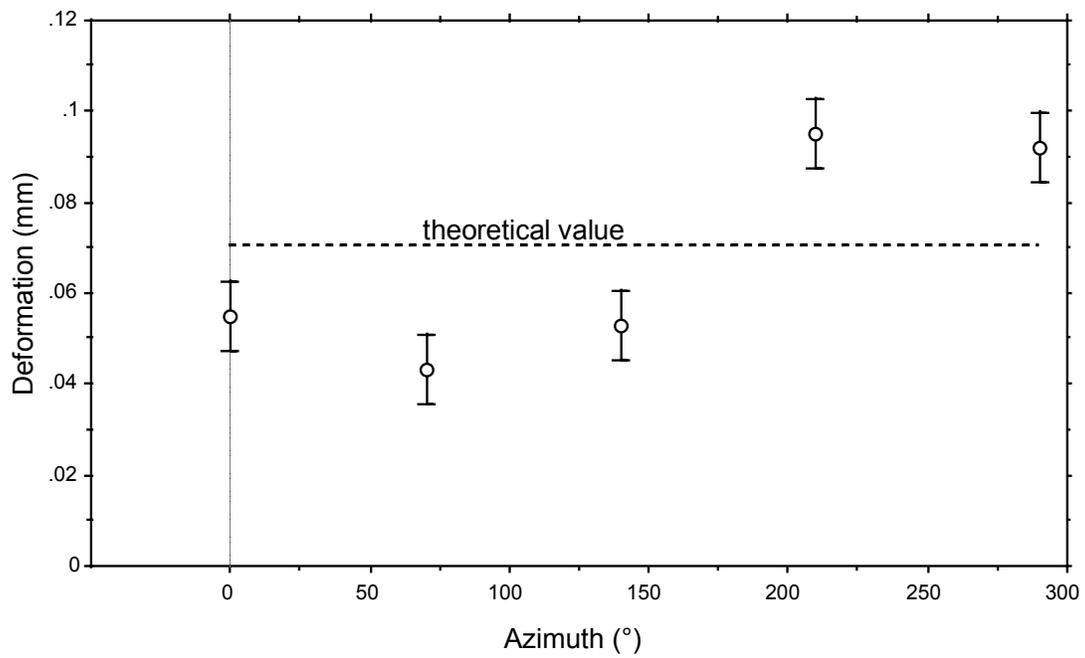

**Figure 4:** Deformation of the edge of the 3.7-m container under 1 mm of mercury

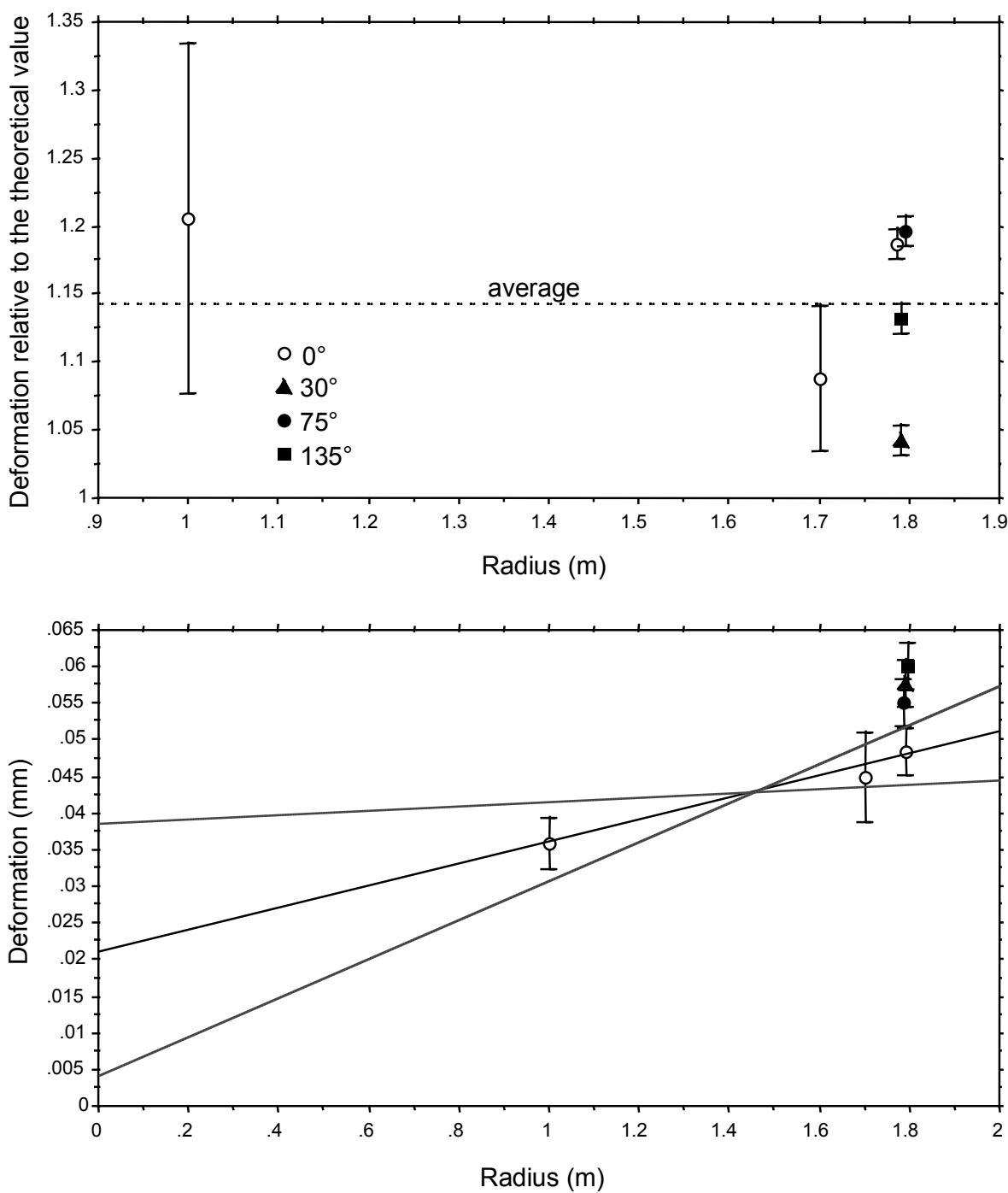

**Figure 5:** *Top:* Deformation of the edge of the 3.7-m container under 2 bottles of mercury at equal distance from the center each at one end of a diameter; measurements along different azimuths and radiuses are shown; units are relative to the calculated values from computer simulations. *Bottom:* Residual deformation of the edge of the container under a bottle of mercury at the specified radius and azimuth (same as top graph), the theoretical value being substracted. The lines are the slope through the points of the 0 degree azimuth and the limits of the 95% confidence interval.

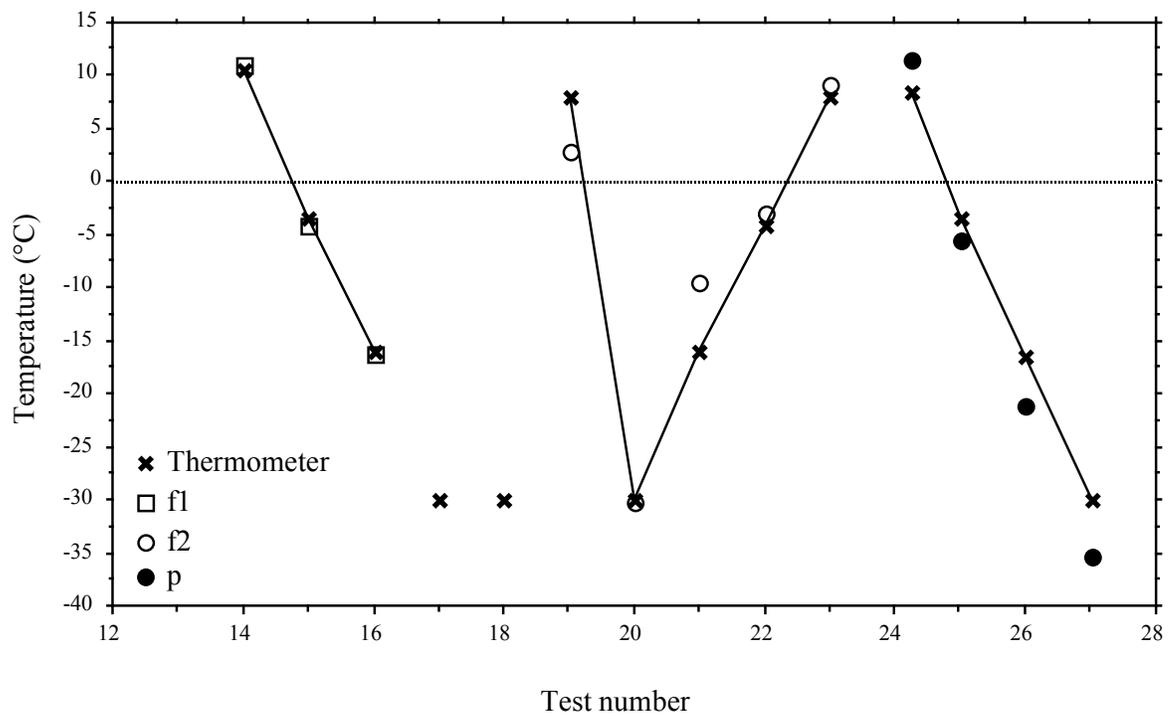

**Figure 6:** Temperature calculated from the total deformation of the 1-m flat mirror for the tests where a steel base was used. Typical standard errors are given by the spot half width.

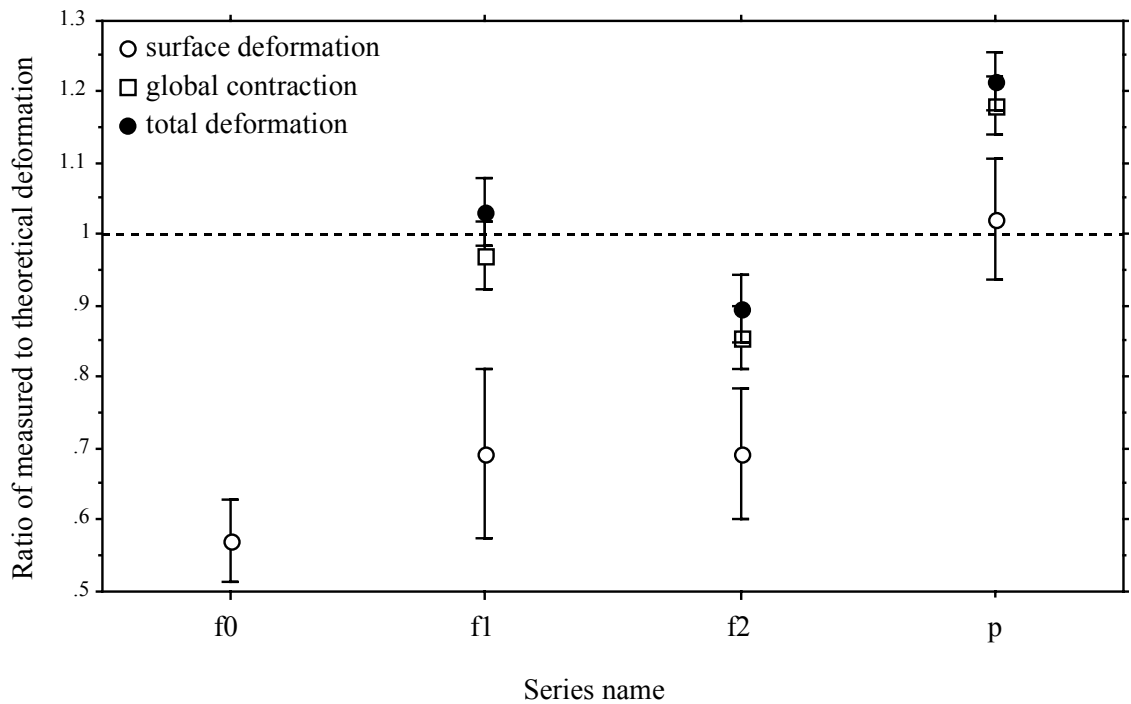

**Figure 7:** Ratio of the average measured deformations to the theoretical deformations of the 1-m flat mirror for all series. Error bars are standard errors.

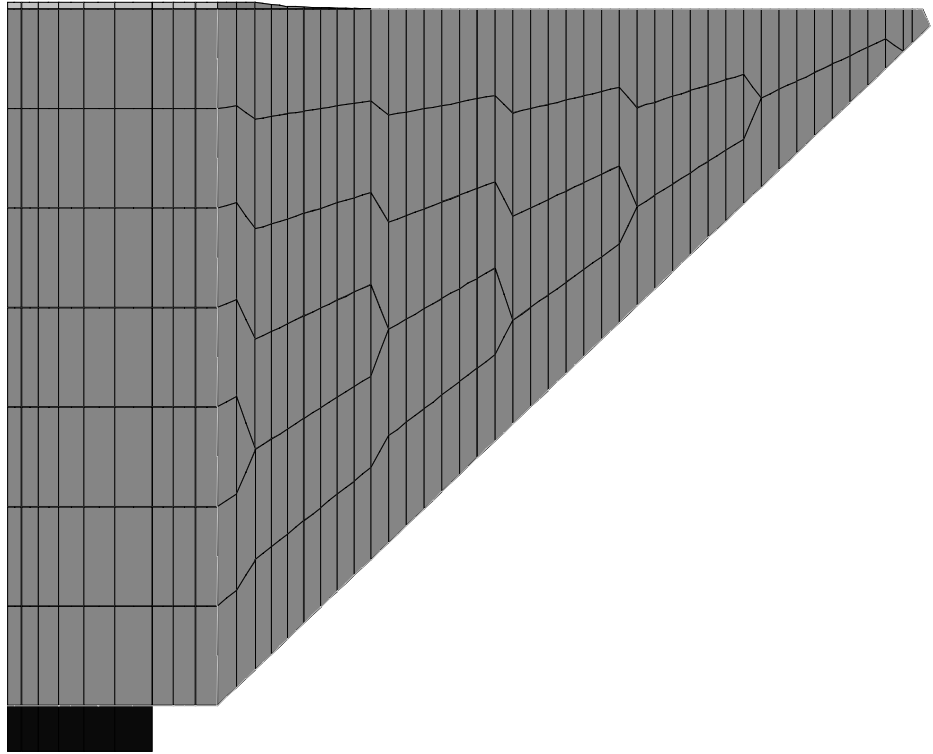

**Figure 8:** Model of the 1-m flat mirror used to study temperature changes

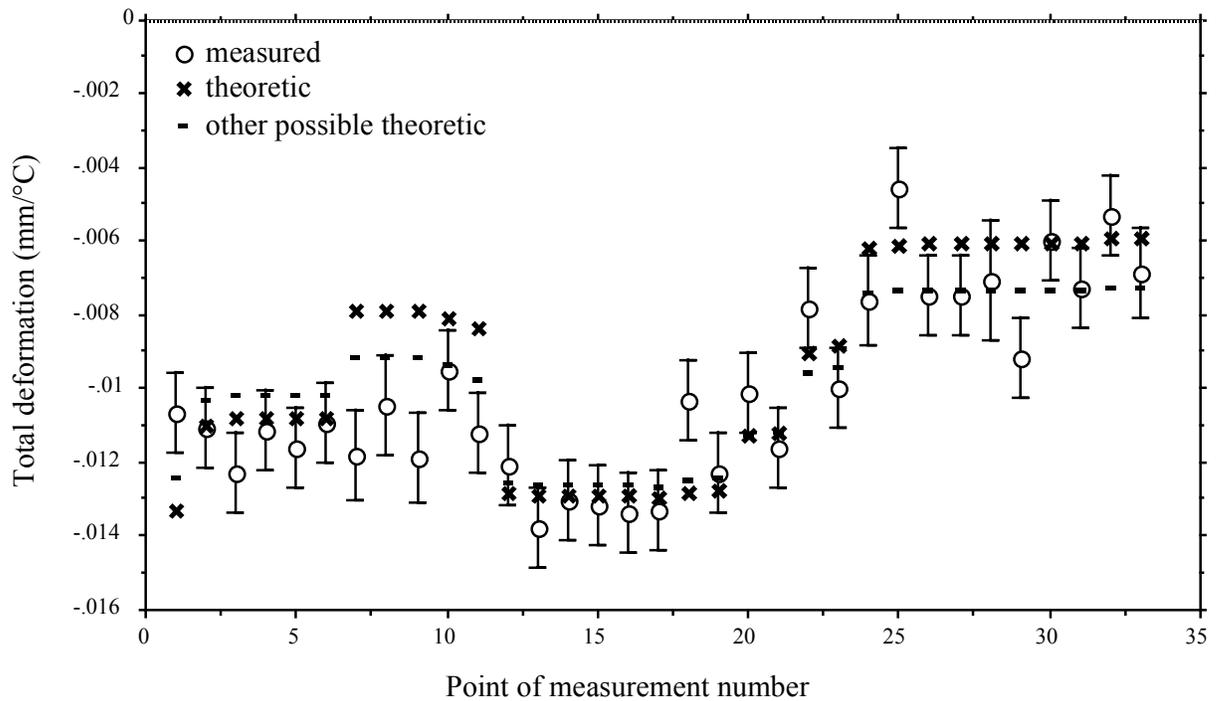

**Figure 9:**    Average total deformation of the 1-m flat mirror for a reduction in temperature of 1°C at all 33 points of measurements. The error bars of the measured deformation are standard errors. The dashes represent the result of a theoretical calculation from computer simulations with 2 modifications to the model: a) the 4.5 cm top of the cylinder has been made conical with the top surface radius 1 cm smaller than at the center of the cylinder; b) the coefficient of thermal expansion of the Kevlar has been increased by 8x10⁻⁶/°C. The position numbers are ordered by the distance to the mirror center.

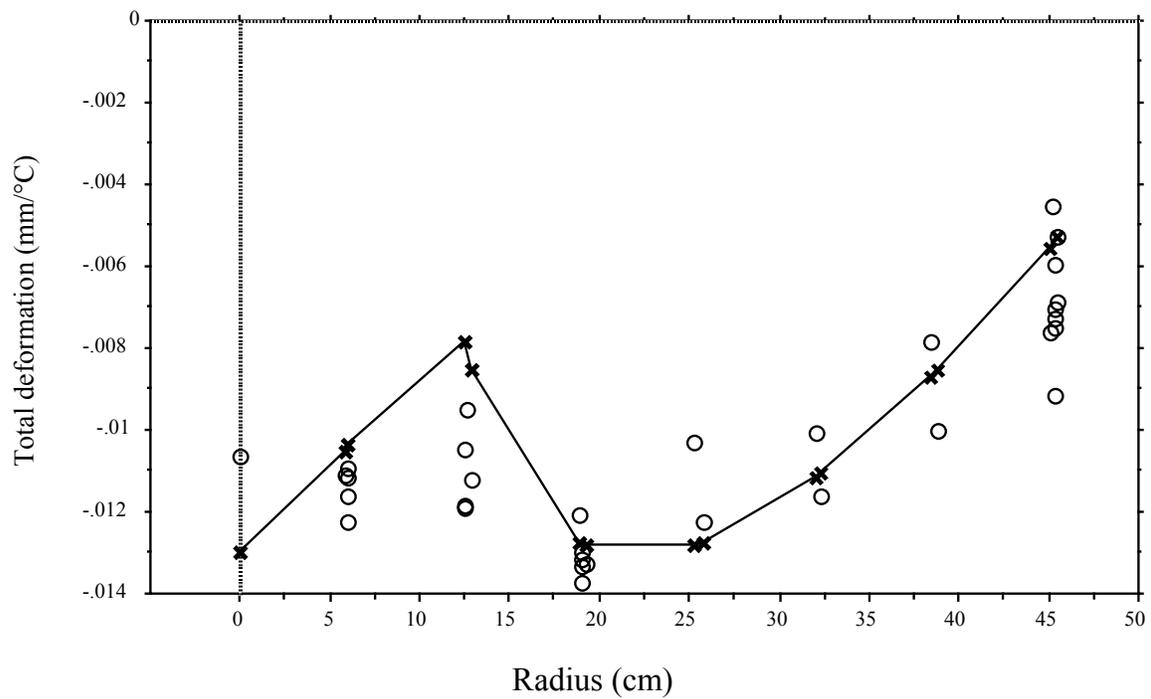

**Figure 10:** Average total deformation of the 1-m flat mirror for a reduction in temperature of 1°C at all 33 points of measurements

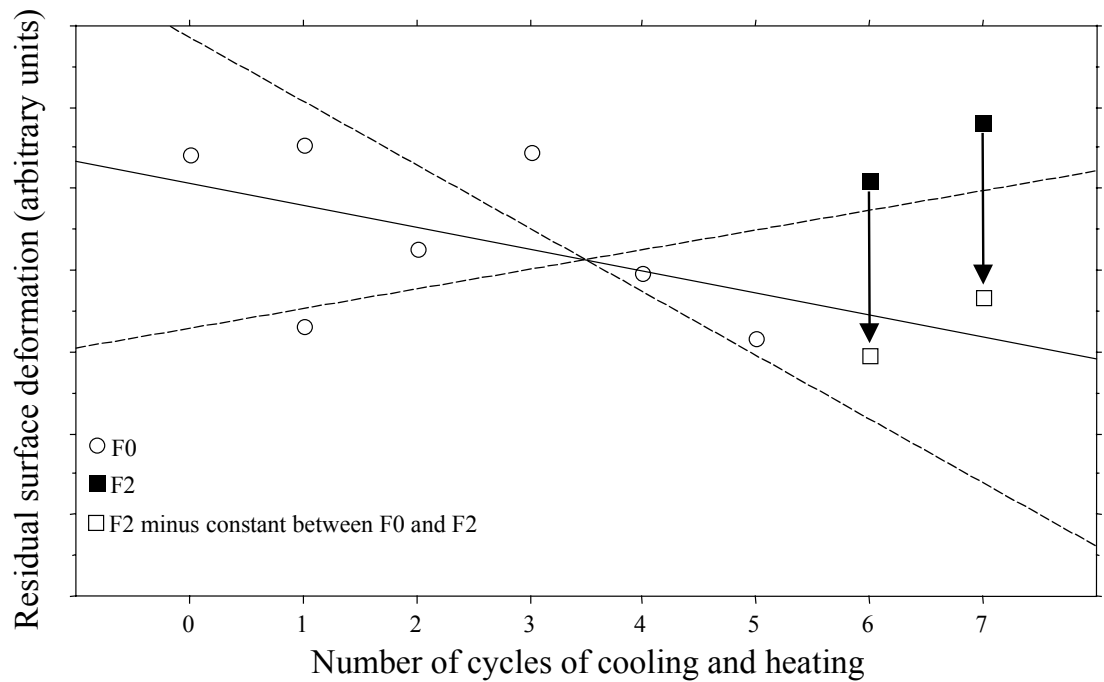

**Figure 11:** Difference between the calculated and measured deformations at the end of each cycle of cooling then heating. Because the reference at 0°C is different for each series, a constant modelling the difference between the series F0 and F2 was added to the linear model. The dotted lines show the limits of the 95% confidence interval of the slope.